\documentclass{ws-procs9x6}

\usepackage{amssymb}
\usepackage{graphicx}
\usepackage{dcolumn}
\usepackage{bm}
\usepackage{amsfonts}
\usepackage{latexsym}
\usepackage{amsmath}
\usepackage{psfrag}
\usepackage[dvips]{epsfig}

\newcommand{\be}{\begin{equation}}
\newcommand{\ee}{\end{equation}}
\newcommand{\bea}{\begin{eqnarray}}
\newcommand{\eea}{\end{eqnarray}}
\newcommand{\bw}{\begin{widetext}}
\newcommand{\ew}{\end{widetext}}

\begin{document}

\title{Radiation reaction in curved space-time: local method.\footnote{\uppercase{T}his work was
supported in part by the \uppercase{RFBR} grant 02-04-16949. }}

\author{Dmitri Gal'tsov,  Pavel
Spirin and Simona  Staub }

\address{Department of Theoretical Physics,
Moscow State University, 119899, Moscow, Russia \\
E-mail: galtsov@phys.msu.ru}

\maketitle

\abstracts{Although  consensus seems to exist about the validity
of equations accounting for radiation reaction in curved
space-time, their previous derivations were criticized recently as
not fully satisfactory: some ambiguities were noticed in the
procedure of integration of the field momentum  over the tube
surrounding the world-line. To avoid these problems we suggest a
purely local derivation dealing with the field quantities defined
only {\em on} the world-line. We consider point particle
interacting with scalar, vector (electromagnetic) and linearized
gravitational fields in the (generally non-vacuum) curved
space-time. To properly renormalize the self-action in the
gravitational case, we use a manifestly
reparameterization-invariant formulation of the theory. Scalar and
vector divergences are shown to cancel for a certain ratio of the
corresponding charges. We also report on a modest progress in
extending the results  for the gravitational radiation reaction to
the case of non-vacuum background.}

\section{Introduction}
Study of the radiation reaction problem in classical
electrodynamics initiated by Lorentz and Abraham by the end of the
19-th century~\cite{Ab05}, remained an area of active research
during the whole 20-th century. Although with the development of
quantum electrodynamics this problem became somewhat academic, it
still attracts attention in connection with new applications and
new ideas in fundamental theory. Current understanding of the
radiation reaction has emerged in the classical works of
Dirac~\cite{Di38}, Ivanenko and Sokolov~\cite{IvSo48},
Rohrlich~\cite{Ro65}, Teitelboim~\cite{Te70} and some other (for a
recent discussion see~\cite{galpav}).

The Lorentz-Dirac equation was covariantly generalized to curved
space-time with arbitrary metric by DeWitt and Brehme in 1960. In
their paper (\cite{DeWitt:1960fc}) an elegant technique of
covariant expansion of two-point tensor quantities was introduced
which later became a basis  of the perturbative quantum field
theory calculations in curved space-time. It was used for
calculation of the field momentum within a small tube surrounding
the world-line of a point charge which resulted in the charge
equation of motion with the radiation damping term. The main
difference with the flat space case was the presence of the tail
term depending on the entire history of a charge. Its presence
signals  violation of the Huygens' principle in curved space:  a
sharp pulse of light does not in general remain sharp, but
gradually develops a "tail".  The equation (as extended by Hobbs
to non-vacuum metrics) \cite{hobbs}) reads
\begin{eqnarray}\label{hobb}
m \ddot z^\alpha = && e F^{\text{in }\alpha }_{\;\;\;\;\;\beta }
\dot z^\beta +   \frac{2}{3}e^2(\dddot z^{\alpha}-\ddot z^{2}\dot
z^{\alpha })\nonumber\\ + &&\frac{e^2}{3}  \biggl(
R^{\alpha}_{\;\;\beta } \dot z^\beta   + R_{\gamma\beta } \dot
z^\gamma \dot z^\beta \dot z^\alpha  \biggr)  +  e^2\, \dot
z^\beta \int_{-\infty }^{\tau } f^{\alpha}_{\;\;\beta \gamma }
\dot z^{\gamma }(\tau')d\tau '.
\end{eqnarray}
where $F^{\text{in }\alpha }_{\;\quad\beta }$ is the incoming
electromagnetic field $R^{\alpha}_{\;\;\beta }$ is the Ricci
tensor and $f^{\alpha}_{\;\;\beta \gamma } $ is  some two-point
function taken on the world-line. This results was later
generalized to other massless fields --- scalar and linearized
gravitational. The gravitational radiation reaction was discussed
long ago by C. DeWitt-Morette and Jing (\cite{morette}), and more
recently reconsidered in detail by Mino, Sasaki and Tanaka
(\cite{mino}) for vacuum background metrics. In \cite{mino} an
extension of DeWitt-Brehme method was used consisting in
integration of the gravitational field momentum flux across the
small world-tube surrounding the particle world-line. The equation
of motion obtained in (\cite{mino}) can be presented as
\begin{align}
\ddot z^{\mu }=-\frac{1}{2} (g^{\mu \nu }+ \dot z^{\mu } \dot
z^{\nu })  (2 h_{\nu \lambda \rho }^{\text{ tail}} - h_{\lambda
\rho \nu }^{\text{ tail}}) \dot z^{\lambda }  \dot z^{\rho }.
\end{align}
where (contrary to the case of (\ref{hobb})) the motion is
supposed to be geodesic, so the local higher-derivative terms
vanishes within the linear approximation, and the reaction force
is given entirely by the tail term
\begin{align}
h_{\mu \nu \lambda }^{\text{ tail}}=4mG \int_{-\infty }^{\tau
^{-}}\biggl( G^{\text{ ret}}_{\mu \nu \tau \sigma ;\lambda }
-\frac{1}{2}g_{\mu \nu } G^{\rho \text{ ret}}_{\;\;\rho  \nu \tau
\sigma ;\lambda } \biggr) z(\tau ) ,z(\tau' )\dot z ^{ \tau }
(\tau ') \dot z ^{ \sigma }(\tau ').
\end{align}

Although there is consensus about the {\em validity} of these
equations, their previous {\em derivation} is   somewhat
problematic.  In fact, as was discussed  recently \cite{SaPo}, all
calculations involving integration of the field momentum located
the small tube surrounding the particle world-line contain some
yet unsolved problems.  One such problem consists in computing the
contributions of "caps" at the ends of the chosen tube segment:
contributions of "caps" were rather {\em conjectured} than
rigorously calculated. Another problem constitutes the singular
integral over the internal boundary of the tube which was simply
discarded in these calculations with no clear justification. In
addition, the usual mass-renormalization is not directly
applicable in the gravitational case since, due to the equivalence
principle, the mass does not enter at all into the geodesic
equations (in \cite{mino} the mass parameter was actually
reintroduced by hand).

Several new derivations were suggested during past few years. One
is based on the redefinition of the Green's functions of massless
fields in curved space-time proposed  by Detweiler and Whiting
\cite{Detweiler:2002mi}. But this scheme involves an axiomatic
assumption about the nature of the singular term and so it does
not help to solve the above problems. Another scheme was suggested
by Quinn and Wald \cite{quwa} under the name of an "axiomatic
approach to radiation reaction". This scheme makes use of some
intuitive "comparison axioms", which do not follow from the first
principles either.  A recent attempt by Sanchez and Poisson
\cite{SaPo} to overcome the above difficulties  in fact is tight
to a particular model of an extended particle (a "dumbbell"
model). Thus, though the results obtained within several different
approaches agree in the final form of non-divergent terms, their
consistent derivation {\em from the first principles} is still
lacking. Also, the derivation of the gravitation radiation
reaction force for non-vacuum background metrics remains an open
problem.

Here we briefly report on the new derivation (more detailed
version will be published elsewhere) of the radiation reaction in
curved space-time which has an advantage to deal with the fields
{\em only on the world-line} and not to involve the volume
integrals over the world-tube at all. This provides a more
economic calculation and at the same time removes  objections
raised in \cite{SaPo}. The problem of divergences is relocated to
the definition of the delta function with the support lying on the
boundary of the integration domain, but this is exactly the same
problem which is encountered in the flat space case where the
approbated prescription amounts to the point-splitting procedure.
Thus the local method is free from ambiguities which were met in
the previous curved space calculations and can be considered as an
adequate solution of the radiation reaction problem in curved
space-time from the first principles. In general, our final
results are conformal with those derived previously; in addition
we partly remove the restriction by vacuum metrics in the
gravitational case. Our signature is ($-$ + + +).

\section{Non-geodesic motion} Here we consider the non-geodesic motion
of point particle along the affinely parametrized world-line
$x^\mu=z^\mu(\tau)$ interacting with the scalar $\phi$ and vector
$A^\mu$ fields. The total  action is the sum $S=S_p+S_f$, where
the world-line part reads
\begin{align}\label{sp}
S_{\rm p}=- m_{0}\int  (1+q\phi)\sqrt{-\dot{z}^2}d\tau +e\int
A^{\mu}\dot{z}_{\mu}d\tau,
\end{align}
($m_0$ being the bare mass, $q,\,e$ -- the scalar and electric
charges) while the volume part is
\begin{align}\label{sf}
S_{\rm f}=-\frac{1}{4\pi}\int \left( \frac{1}{2}\partial_{\mu}\phi
\partial^{\mu}\phi+\frac{1}{4} F^2 \right)\sqrt{-g}d^4 x.
\end{align}

Our local approach to radiation reaction simply consists in the
substitution of the proper fields into the particle equation of
motion (for brevity we omit the external force)
\begin{equation}\label{peq}
 m_0(1+q\phi)\ddot{z}^{\mu}=eF^{\mu}_{\;\;\nu}\dot{z}^{\nu}-m_0
q\Pi^{\mu\nu}\phi_{;\,\nu}
\end{equation}
where the velocity-transverse projector in the gauge
$\dot{z}^2=-1$  reads
$$\Pi^{\mu\nu}=g^{\mu\nu}+\dot{z}^{\mu}\dot{z}^{\nu}.$$

The equations for the scalar field and the 4-potential ($F=dA$)
read
\begin{align}\label{emotions}
& \Box \phi =4\pi \rho   \\
& \Box A_{\mu }-R_{\mu }^{\nu }A_{\nu }= -4\pi j^{\,\mu}
\end{align}
where  the covariant D'Alembert operator is understood
$\Box=D_{\mu}D^{\mu}$, and the source terms are standard. The
Green's functions are defined in the usual way starting with the
Hadamard solution (in the scalar case):
\begin{equation}\label {hadamard}
G^{(1)}(x,z)= \frac{1}{(2\pi  )^2} \left[
\frac{\Delta^{1/2}}{\sigma }+ v \ln  \sigma +w \right],
\end{equation}
where $\sigma(x,z)$ is the Synge's world  function, $\Delta$ is
van Vleck determinant, and $v, w$ satisfy the system
\begin{align}
& \Box v=0,  \nonumber \\ & 2v+(2v_{;\mu}-v \Delta
^{-1}\Delta_{;\mu }) \sigma^{;\mu} + \Box \Delta^{1/2} + \sigma
\Box w =0. \nonumber
\end{align}
In the vector case
\begin{equation}  \label{luthor}
G^{(1)}_{\;\;\mu \alpha }= \frac{1}{(2\pi)^2} \left ( \frac{u_{\mu
\alpha } }{\sigma }  + v_{\mu \alpha }\text{ ln } \sigma  + w_{\mu
\alpha }\right ),
\end{equation}
where $ u_{\mu \alpha }(x,z)$, $ v_{\mu \alpha }(x,z) $ and
$w_{\mu \alpha } (x,z)$ are bi-vectors, satisfying a similar
system. In particular, $ u_{\mu\alpha }=\bar g_{\mu \alpha }\Delta
^{1/2} $, where $\bar g_{\mu \alpha }$ is the bi-vector of
parallel transport. Notation is that of DeWitt and Brehme: indices
$\alpha,\beta,...$ are associated with the ``emission'' point $z$,
while $\mu,\nu,...$ --- with the ``observation'' point $x$. When
both points are taken on the world-line, we use the first set to
denote $z^\alpha(\tau')$ associated with the integration variable
$\tau'$, and the second set to denote $z^\mu(\tau)$, where $\tau$
is the proper time in the resulting equation.
 In terms of these quantities the retarded solutions of the field
equation read
\begin{align} \label{scagree}
\phi^{\text {ret}}(x)= m_0 q \int \limits_{-\infty}^{\tau_{\rm
ret}(x)}\left[-u \delta (\sigma )+v \theta (-\sigma
)\right]d\tau',
\end{align}
\begin{equation} \label{emintens}
F^{\rm ret}_{\mu\nu}(x)=-2e \int \limits_{-\infty}^{\tau_{\rm
ret}(x)} \left[u_{[\mu \alpha} \sigma_{\nu]} \delta '(\sigma )  +
  (u_{[\mu \alpha; \,\nu]} +v_{[\mu \alpha}\sigma_{\nu]}) \delta
(\sigma ) + v_{[\nu \alpha ; \,
\mu]}\right]\dot{z}^{\alpha}d\tau',
\end{equation}
where an anti-symmetrization over the indices $\mu$ and $\nu$ is
used.

 When substituted into the equations of motion, two-point
functions become localized on the world-line, so we are led to use
the expansions of the Synge's function  $\sigma (z(\tau),z(\tau' ))
$(in terms containing delta-function and its derivative) as follows
\begin{equation}
 \sigma (z(\tau),z(\tau' ))=
 \sum_{k=0}^{\infty} \frac{1}{k!}\frac{D^k}{d\tau^k}
 \sigma (\tau,\tau) (\tau -\tau')^k
\end{equation}
In what follows we will  denote by dots the quantities
\begin{align}
\dot \sigma = \sigma _{\alpha} \dot z^\alpha
\;\;\;\;,\;\;\;\;\ddot \sigma = \sigma _{\alpha\beta } \dot
z^\alpha \dot z^\beta +
  \sigma _{\alpha}\ddot z^\alpha,\quad {\rm etc.}
\end{align}
Denoting   the difference as $s=\tau -\tau'$, we get for $\sigma$
an expansion similar to that in the flat space
\begin{equation}
\sigma (s)= - \frac{s ^2}{2}- \ddot z^2(\tau) \frac{s
^4}{24}+\mathcal{O}(s^5),
\end{equation}
but with dots corresponding to covariant derivatives along the
world-line. Similarly, for the  derivative of $\sigma$ with
respect to $z^\mu (\tau)$ one finds
\begin{equation}\label{signu}
\sigma ^{\mu}(s)=  s \left(\dot z^\mu -\ddot z^{\mu}\frac{s}{2}
+\dddot z^\mu\frac{s^2}{6} \right)+\mathcal{O}(s^4).
\end{equation}
This quantity is a vector at the point $ z(\tau ) $ and a scalar
at the point $ z(\tau ')$ where the index $\mu$ now corresponds to
the point $ z(\tau ) $: $\sigma ^{\mu}=\partial \sigma
(z,z')/\partial z_{\mu}$.

An expansion for the delta-function reads
\begin{align}
\delta(-\sigma)=\delta(s^2/2)+s^4\frac{\ddot z^2(\tau)}{24}
\delta'(s^2/2)+...
\end{align}
and an analogous expansion is easily obtained for its derivative.
Since the most singular term is $u\delta' (\sigma ) \sigma_{\nu }$
the maximal expansion order to be retained is $s^3$. This also
means that we have to retain for the delta-function only the
leading term: $\delta(\sigma)=\delta(s^2 /2)+\mathcal{O}(s^4)$.
The delta-function of the squared argument can be regularized in
exactly the same way as in the flat space as
$$\theta (s) \delta(s ^2 /2) \to \theta (s) \delta([s
^2-\varepsilon^2] /2)=\delta(s-\varepsilon)/\varepsilon,$$ where
the positive regularization parameter $\varepsilon\to 0$ of the
dimension of length is introduced.  One finds in particular
  $$\theta (s)
\delta'(\sigma) \to \theta (s)\frac{1}{s}\frac{d}{d s}
\delta\left(\frac{s
^2-\varepsilon^2}{2}\right)=\frac{1}{\varepsilon
s}\delta'(s-\varepsilon)$$. We have to perform the expansions up
to the third order in $s$ in all terms containing delta-function
and its derivative (local terms). For the function $u$ we will
have Ricci-terms involved into expansions, up to third order one
finds:
\begin{align}\label{utochka}
u= 1+\frac{1}{12}R_{\sigma  \tau }\dot z^\tau \dot z^\sigma s ^2 ,
\quad u_{;\nu }= \frac{1}{6} R_{\nu \tau }\dot z^\tau s,
\end{align}
and similarly in the vector case
\begin{align}\label{s6}
u_{\nu \alpha }\sigma ^{;\mu }\dot z^{\nu }  \dot z^{\alpha }  =
-s \dot z^\mu  + \frac{s^2}{2} \ddot z^\mu -s^3 \left( \frac{1}{6}
\dddot z^\mu +\frac{1}{12}R_{\lambda \nu } \dot z^\lambda \dot
z^\nu \dot z^\mu +\frac{1}{2} \ddot {z}^2 \dot z^\mu \right).
\end{align}
The functions of $v$-type in the tail term can not be found in a
closed form.

Collecting all the contributions,  one obtains the following
expressions for the scalar and  electromagnetic reaction forces:
\begin{align}\label{scfin}
f^{\mu}_{\rm sc}&=  m^2 q^2\left[\frac {\ddot z^{\mu}}{2 \epsilon
}+ \Pi^{\mu\nu}\left(\frac {1}{3}\dddot z_{\nu}+\frac {1}{6}
R_{\nu \tau }\dot z^\tau - \int \limits_{-\infty }^{\tau }
v_{;\,\nu}\, d\tau '\right)\right.
  \left.-\ddot z^{\mu}\int
\limits_{-\infty }^{\tau } v d\tau'\right],\\
f^{\mu}_{\rm em}&=
e^2\left[-\frac{\ddot{z}^{\mu}}{2\epsilon}+\Pi^{\mu\nu}
\left(\frac{2}{3}\dddot z_{\nu}+\frac{1}{3}R_{\nu\alpha}\dot
z^\alpha\right) +  2 \dot z^{\nu }(\tau )\int \limits_{-\infty
}^{\tau }v^{[\mu }_{\,\,\,\alpha;\nu]}  \dot z^{\alpha } (\tau'
)d\tau'\right].
\end{align}
Divergent terms can be absorbed by the renormalization of mass.
\begin{align}\label{cancell}
m=m_0+\frac{1}{2\epsilon}(m_0^2 q^2-e^2)
\end{align}
Note different signs of the divergent terms for scalar and vector
self-forces, so for a "BPS" particle with the ratio of charges $m
|q|=|e|$  the model is free of divergencies. Finite contributions
coincide with the previous results obtained via (somewhat
questionable) world-tube derivations. Thus, in view of the
criticisms in \cite{SaPo},  our derivation may be considered as
confirmation of these results. Technically, the local calculation
is substantially simpler than the world-tube one.

\section{Neutral particle at geodesic motion}

In the case of gravitational radiation reaction we have the only
one parameter ---  particle mass, which actually does not enter
into the geodesic equation.  Thus it is not possible to use the
above renormalization scheme. This problem can be remedied using a
manifestly reparametrization invariant treatment. This is done by
introducing the einbein  $e(\tau )$ on the world line acting as a
Lagrange multiplier:
\begin{align}\label {act}
S[ z^\mu, e]=-\frac{1}{2}\int \left[ e (\tau )\, g_{\mu \nu } \dot
z^{\mu } \dot z^{\nu } +\frac{m^2}{e(\tau )}\right]d\tau ,
\end{align}

Varying (\ref {act}) with respect to $z^\mu (\lambda )$ and
$e(\tau )$ gives
\begin{align}\label{mot}
\frac{D}{d\tau }(e\dot z^\mu )=0,\quad e=\frac{m}{\sqrt{ -g_{\mu
\nu } \dot z^\mu \dot z^\nu } },
\end{align}
and we obtain the geodesic equation in a manifestly
reparametrization invariant form
\begin{align}\label{caf}
\frac{D}{d\tau }\left(\frac{\dot z^\lambda}{\sqrt{ - g_{\mu \nu }
\dot z^\mu \dot z^\nu }}\right)=0.
\end{align}
We split the total metric into background and perturbation due to
point particle $g_{\mu \nu } \to \hat{g}_{\mu \nu }=g_{\mu \nu }
+\kappa h_{\mu \nu }$. Assuming now that the particle motion with
no account for radiation reaction is geodesic on the background
metric, the perturbed equation in the leading order in $\kappa$
will read
\begin{align}\label{prop}
\ddot z^\mu = \frac{\kappa }{2}\left(g^{\mu \nu }-\frac{\dot z^\mu
\dot z^\nu}{\dot z^2}\right) ( h_{\lambda \rho ;\nu}-2h_{\nu
\lambda ;\rho}) \dot z^\lambda \dot z^\rho,
\end{align}
where  contractions are with the background metric.

The derivation of the field equations for the metric perturbation
in the general case of non-vacuum background is non-trivial
\cite{sciama}. It is expected that the particle stress-tensor
$\overset {(P)}{T^{\mu \nu }}$ has to be divergence-free with
respect to the background metric, this allows one to derive the
equation for the metric perturbation in a harmonic gauge, which
leads to a manageable Green's function. But expanding the Bianchi
identities for the full metric one finds that this is only
possible if the Einstein tensor $G_{\mu\nu}$ and metric
perturbation $h_{\mu\nu}$ satisfies an additional equation
\begin{align}\label{cons}
G^{\lambda}_{ \mu}h_{ \, ; \,
\lambda}-G_{\lambda\rho}h^{\lambda\rho \, ; \,
\mu}=\mathcal{O}(\ddot{z}^{\mu}).
\end{align}
Otherwise, the ``naive'' particle stress tensor on a given
background is not enough, and  construction of a reliable source
term for metric perturbation becomes a complicated problem.
Clearly, there is no general solution  to the constraint
(\ref{cons}), but some particular cases can be found. One can
notice  that this equation is identically satisfied for Einstein
metrics $R_{\mu\nu}=-\Lambda g_{\mu\nu}$. Another case is
conformally-flat metrics with some special conformal factor
(details will be given elsewhere). Provided the Eq. (\ref{cons})
holds,  one can derive (using the results of Sciama et al.
\cite{sciama}) the following equation for the trace-reversed
perturbation $\psi_{\mu
\nu}=h_{\mu\nu}-g_{\mu\nu}h^\lambda_\lambda/2$:
\begin{align}\label{firstier}
\psi^{\mu \nu ;\xi }_{\quad\;\;;\xi }+ 2 R^{\,\mu \;\nu
}_{\;\,\xi\;\,\rho }\,\psi^{\xi \rho }-2\psi ^{(\mu }_{\;\;\sigma
}R^{\nu )\sigma }+ \psi^{\mu \nu }R - g^{\mu \nu }R_{\alpha \beta
}\psi^{\alpha \beta}= -2 \kappa^2 \overset {(P)}{T^{\mu \nu }}.
\end{align}
The corresponding Hadamard's function is similar to (\ref{luthor})
with four-index bi-tensors $u _{\lambda\rho \alpha \beta },\; v
_{\lambda\rho \alpha \beta }$.  The retarded solutions is
constructed in an analogous way and substituted into the equation
of motion (\ref{prop}) leading to an integral involving bi-tensor
quantities depending on two points on the world-line
$\tau,\;\tau'$. As before, one can distinguish  four different
contributions: terms proportional  to the delta function,   the
derivative of the delta function,  the derivative of the Heaviside
function, and the tail term. Performing  series expansions in the
first three cases up to the third order in  $\tau -\tau '=s$  one
finds a local contribution to the self-force. To facilitate
expansions of bi-tensors one has first to reduce them to scalars
at the point of expansion, e.g. quantities like $u_{\nu \lambda
\alpha \beta } z^ \alpha (\tau ') z^ \beta (\tau ')$ behave   as a
scalar with respect to  the point $z(\tau ')$ and tensors with
respect to $z(\tau )$. Then expansion in $s$ around $\tau$ is
straightforward.

Contrary to the non-geodesic case, now we have to drop all
(covariant) derivatives with respect to the proper time of the
second and higher order, with the only exception of divergent
term. Actually, the divergent term is proportional to the
acceleration, so formally it vanishes for the geodesic motion. But
still it may be argued that its infinite value demands some
renormalization to be made, and we have to find an appropriate
quantity to be  renormalized . Since the mass does not enter the
equation, we invoke the einbein $e(\tau)$, and insert as the
corresponding background quantity $e_0=$const. Collecting the
expansions we obtain
\begin{align}
\ddot z^\mu = \kappa^2 e_0 \left[ \frac{7}{2\epsilon } \ddot z^\mu
- \frac{11}{6} \Pi^{\mu\nu} R_{\nu \lambda }\dot z^\lambda   +
F^{\mu}_{\rm tail}\right].
\end{align}
Recall, that Ricci term is not arbitrary here, but has to be
proportional to the metric for consistency. So actually the only
local term here is proportional to the four-velocity $\dot z_\nu$
and vanishes by virtue of the projector. The tail term is
\begin{align}
F^{\mu}_{\rm tail} =\frac{\kappa^2\,e_0}{4} \Pi^{\mu\nu}\int
\limits_{-\infty }^\tau [4 v _{\nu \lambda\alpha \beta ;\rho}-&2(
g_{\nu \lambda } v^\sigma_{\sigma \alpha \beta ;\rho}+
v_{\lambda\rho \alpha \beta ;\nu})- \nonumber\\-&g_{\lambda \rho }
v^\sigma_{\sigma \alpha \beta ;\nu})] \dot z'^{\alpha }\dot
z'^{\beta } \dot z^\lambda \dot z^\rho d\tau'.
\end{align}
Renormalization of the einbein is performed as
\begin{equation} \label{renorm}
\biggl( \frac{1}{e_0}-\frac{7\kappa ^2}{2\epsilon } \biggr) \ddot
z^\mu   = \frac {1}{e}\ddot z^\mu.
\end{equation}
Finally the choice of the affine parameter $\dot z^2=-1$,
equivalent to setting $e=m$, leads to the final form
\begin{align} \label{antifin}
& \ddot z =m\kappa^2 \Pi^{\mu\nu}  \left[ -\int _{-\infty }^\tau [
\left( 2v _{\nu \lambda\alpha \beta  ;\rho}-
 g_{\nu \lambda }g^{\sigma
\tau }v _{\sigma \tau \alpha \beta ;\rho} \right ) - \right.\nonumber \\
- &\left. \left( v _{\lambda\rho \alpha \beta ;\nu}-1/2 \,
g_{\lambda \rho }g^{\sigma \tau }v_{\sigma \tau \alpha \beta
;\nu}\right)] \dot z'^{\alpha }\dot z'^{\beta }d\tau ' \dot
z^\lambda \dot z^\rho  \right].
\end{align}
   This coincides with the result obtained in
\cite{mino,quwa}. We have thus shown that this equation remains
valid for a class on non-vacuum metrics, in particular, for
Einstein spaces.

\section{Discussion}
We have presented a local calculation of radiation reaction in
both geodesic and non-geodesic cases which do not involve any
ambiguous integrals outside the world-line and thus is free from
the corresponding problems. It is based on the expansion of
bi-tensor quantities only on the world line and technically  is
much simpler than the DeWitt-Brehme type approach. We have also
generalized the gravitational radiation reaction to some
non-vacuum metrics, however, satisfying  an additional condition.
The general non-vacuum case can not be treated within the test
body/external field approximation because the global Bianchi
identity demands to take into account the perturbation of the
background matter as well.

\section*{Acknowledgments}
DG is grateful to the Department of Physics of the National
Central University (Taiwan) for hospitality and support during the
Conference. He would like to thank J. Nester and C.-M. Chen for
useful discussions. The work was supported in part by the RFBR
grant 02-04-16949.

\end{document}